\documentclass[
reprint,
amsmath,
amssymb,
 aps,nofootinbib]{revtex4-2}

\usepackage{graphicx}
\usepackage{dcolumn}
\usepackage{bm}
\usepackage{hyperref}
\usepackage{orcidlink}

\usepackage{soul}
\usepackage{braket}
\usepackage[dvipsnames]{xcolor}
\usepackage{siunitx} %
\usepackage{subcaption}
\usepackage{booktabs}

\newcommand{\bayes}[2]{\mathcal{B}_{\mathrm{#1}}^{\mathrm{#2}}}

\makeatletter\let\selectlanguage\@gobble\makeatother
\begin{document}
\title{Environmental effects vs.\ modified gravity in the LISA massive black hole binary population}
\author{Lorenzo Copparoni}
\affiliation{SISSA, Via Bonomea 265, 34136 Trieste, Italy and INFN Sezione di Trieste}
\affiliation{IFPU - Institute for Fundamental Physics of the Universe, Via Beirut 2, 34014 Trieste, Italy}

\author{Enrico Barausse}
\affiliation{SISSA, Via Bonomea 265, 34136 Trieste, Italy and INFN Sezione di Trieste}
\affiliation{IFPU - Institute for Fundamental Physics of the Universe, Via Beirut 2, 34014 Trieste, Italy}

\begin{abstract}
Gravitational-wave signals from massive black hole binaries observed by LISA can carry imprints of both the astrophysical environment of the source and possible deviations from general relativity. We investigate whether environmental effects leave a detectable imprint on the LISA binary population, and whether they can mimic modified-gravity effects with the same frequency dependence. As representative channels we adopt accretion and viscous migration in a circumbinary disk for the environmental sector, and a time-varying Newton constant $\dot G$ for the modified-gravity sector. All three effects enter the waveform at the same negative post-Newtonian order and are described, at leading order, by a common phase-deformation parameter, which makes them formally degenerate at the single-event level. Combining Fisher-matrix forecasts with a hierarchical nested-sampling analysis of synthetic catalogs from astrophysically motivated population models, we find that, even under extreme astrophysical assumptions---an active fraction of $50\%$, together with a super-Eddington accretion tail---the population-level posteriors remain fully compatible with vacuum. However, a hierarchical population-wide analysis may yield a non-trivial upper limit on the active fraction and a mild lower bound on the slope of the Eddington-ratio distribution. Environmental effects are therefore unlikely to bias LISA's tests of general relativity with massive black hole binaries in astrophysically realistic scenarios.
\end{abstract}

\maketitle

\section{Introduction}
\label{sec:introduction}

The direct detection of gravitational waves (GWs) from compact-binary coalescences~\cite{LIGOScientific:2016aoc} has turned GW astronomy into a powerful probe of both astrophysics and fundamental physics. 
On the astrophysical side, the growing catalog of events from the LIGO-Virgo-KAGRA (LVK) collaboration~\cite{KAGRA:2021vkt, LIGOScientific:2025slb} has yielded a wealth of information on the formation channels, evolutionary histories and host environments of stellar-mass compact binaries~\cite{KAGRA:2021duu, LIGOScientific:2016vpg, LIGOScientific:2020ufj}. On the fundamental physics side, the observed waveforms---spanning the full inspiral, merger and ringdown---can be compared in detail with the predictions of general relativity (GR) in its strong-field, highly-dynamical regime, providing some of the most precise tests of the theory available to date~\cite{LIGOScientific:2020tif, LIGOScientific:2021sio, LIGOScientific:2026qni, LIGOScientific:2026fcf, LIGOScientific:2026wpt}. 
In the next decade, the space-based Laser Interferometer Space Antenna (LISA)~\cite{colpi_lisa_2024, amaro-seoane_astrophysics_2023} will 
extend the frequency range of GW astronomy to the mHz regime, where the expected primary sources are massive black hole binaries (MBHBs) with total masses in the $10^{4}$--$10^{7}\,M_\odot$ range~\cite{klein_science_2016, Sesana:2007sh, Barausse:2012fy, Bonetti:2018tpf, Barausse:2020mdt}, extreme mass-ratio inspirals (EMRIs) of stellar-mass compact objects into central massive black holes (MBHs)~\cite{Babak:2017tow, Berry:2019wgg}, and the early inspiral of stellar-origin black hole binaries (SOBHBs)~\cite{Sesana:2016ljz, Sberna:2022qbn, Chen:2025qyj}. 
At even lower (nHz) frequencies, pulsar timing arrays (PTAs)
have been observing~\cite{NANOGrav:2023gor, EPTA:2023fyk, Tarafdar:2022toa, Reardon:2023gzh, Xu:2023wog} a possible stochastic background of GWs from MBHBs of mass  $\gtrsim 10^8 M_\odot$.

The gravitational waveform of an inspiralling binary is never, in a real astrophysical system, that of a clean two-body system in vacuum. Gas accretion onto the components and dynamical friction/migration torques from the surrounding gas~\cite{Barausse:2007dy, barausse_can_2014, caputo_gravitational-wave_2020,Toubiana:2020drf, Kocsis:2011dr, Yunes:2011ws,  Speri:2022upm, Copparoni:2025jhq, Chen:2025qyj}
can leave imprints on the binary's phase evolution, whose magnitude can become comparable to the radiation-reaction term at low frequencies.
A growing body of work has now shown that such environmental imprints are in principle detectable with LISA across a number of source classes, from intermediate mass-ratio and EMRI systems~\cite{Barausse:2007dy, barausse_can_2014, Kocsis:2011dr, Yunes:2011ws, Speri:2022upm, Copparoni:2025jhq} to SOBHBs~\cite{caputo_gravitational-wave_2020,Toubiana:2020drf}
and stochastic backgrounds~\cite{Chen:2025qyj}. For PTAs, the effects of the surrounding environment (stellar hardening and triple massive black hole interactions) are
expected to be even more significant~\cite{Bonetti:2017lnj}, possibly explaining the flatter-than-expected background observed in the nHz band~\cite{NANOGrav:2023gor, EPTA:2023fyk, Tarafdar:2022toa, Reardon:2023gzh, Xu:2023wog}.

For MBHBs in the LISA band the situation is more nuanced. Order-of-magnitude arguments~\cite{barausse_can_2014} indicate that the bulk of these sources should be only weakly affected (if at all) by their astrophysical environment. It has more recently been pointed out, however, that this conclusion may break down at the lower-mass end of the population, $\lesssim 10^{5}\,M_\odot$~\cite{garg_imprint_2022, Zwick:2026zks}. Moreover, even when they are small in absolute terms, environmental effects can still bias the tests of GR that LISA MBHBs are expected to enable with unprecedented precision. Accretion and gas drag enter the waveform phase at \emph{negative} post-Newtonian (PN) orders, and so do many of the best-motivated modifications of GR. Both classes of effects therefore affect the same low-frequency, early-inspiral portion of the signal. Examples of such GR modifications include a secular variation of Newton's constant~\cite{Yunes:2009bv, tahura_parameterized_2018}, scalar-dipole radiation, and violations of the strong equivalence principle~\cite{Barausse:2016eii}.

In this work we focus on a particularly evident  realization of this overlap: a GR modification and a set of environmental effects that all enter the inspiral phase at exactly the same $-4\,\mathrm{PN}$ order, and are therefore formally indistinguishable on a single source. As the GR-modification channel we adopt a secular drift of the gravitational constant, $\dot G/G$~\cite{Yunes:2009bv, tahura_parameterized_2018}. Current bounds on the fractional drift rate range from $|\dot G/G|\lesssim 10^{-12}$--$10^{-13}\,\mathrm{yr}^{-1}$ in the quasi-static Solar-System regime~\cite{bambi_response_2005} to much weaker limits in the strong-field, dynamical regime accessible to ground-based GW observations of SOBHBs~\cite{yunes_theoretical_2016}. In the environmental sector, we consider the two expected leading effects of a gaseous circumbinary disk. The first is gas accretion onto the components, which slowly increases their masses and, by angular-momentum conservation, induces a $-4\,\mathrm{PN}$ dephasing~\cite{caputo_gravitational-wave_2020}. The second is gas-driven migration, i.e.\ the torque exerted on the binary by the disk, which contributes at the same $-4\,\mathrm{PN}$ order, with the same frequency dependence as accretion but a different overall prefactor. At leading order, all three effects are therefore captured by a single $-4\,\mathrm{PN}$ deformation of the inspiral phase, and the signal of an \textit{individual} MBHB cannot distinguish among them. The two classes of effects, however, behave differently across a population of sources: environmental contributions are expected to vary in magnitude from one binary to another, whereas a putative $\dot G$ drift is universal. A hierarchical analysis of the full \textit{population} of LISA MBHBs can therefore in principle still disentangle the two, in analogy with what has been proposed for EMRIs~\cite{Kejriwal:2025jao}.

To answer these questions we combine single-event Fisher-matrix forecasts with a hierarchical Bayesian analysis~\cite{Toubiana:2021iuw} of synthetic MBHB catalogs drawn from astrophysically motivated population models, spanning both heavy-seed (HS) and light-seed (LS) formation channels~\cite{barausse_implications_2023}. The most informative sources turn out to be the lowest-mass MBHBs at $z\lesssim 1$, which constitute a small subset of any realistic population. Even under extreme astrophysical assumptions---an active fraction of $50\%$ together with a tail of the Eddington-ratio distribution extending well into the super-Eddington regime---the population-level posteriors remain fully compatible with vacuum, although they do yield a non-trivial upper limit on the active fraction and a mild lower bound on the slope of the Eddington-ratio distribution. 

The remainder of this paper is organized as follows. Section~\ref{sec:pn4} reviews the $-4\,\mathrm{PN}$ dephasing induced by $\dot G$ and by accretion and viscous migration, and introduces a common agnostic phase-deformation parameter. Section~\ref{sec:population_catalogs} describes the MBHB population models that we consider, computes the corresponding Fisher-matrix parameter-estimation errors across the mass--redshift plane, and presents per-event parameter-estimation results on specific $10\,\mathrm{yr}$ realizations. Section~\ref{sec:hierarchical_inference} embeds the single-event measurements in a hierarchical Bayesian framework, derives the $\dot G$ constraint, and performs population-level model comparisons between the environmental, vacuum and modified-gravity hypotheses.
Throughout the paper we adopt geometrized units with $G_N=c=1$ (with $G_N$ the present day Newton constant) unless otherwise stated.

\section{Modifications at $-4\,\mathrm{PN}$ order}
\label{sec:pn4}

Rather than introducing a dedicated template for each candidate effect, we adopt the parameterized post-Einsteinian (ppE) framework~\cite{Yunes:2009ke}, in which a single parameter captures the dominant phase deformation at a given PN order. For comparable-mass binaries near merger---the regime probed most precisely by current LVK tests of GR---negative-PN corrections of the kind considered here are subdominant to radiation reaction. However, at the lower orbital frequencies accessible to LISA MBHBs, these corrections become, as discussed in Sec.~\ref{sec:introduction}, a natural target. We specialize the ppE template to the $\dot G$ case in Sec.~\ref{subsec:Gdot}, to accretion and viscous migration in Sec.~\ref{subsec:env}, and present the resulting Fisher-matrix parameter-estimation results across parameter space in Sec.~\ref{subsec:Det_LISA}.
\subsection{Gravitational constant variation}
\label{subsec:Gdot}
We adopt here a phenomenological modified-gravity model in which the gravitational constant is time-dependent, as would be the case, e.g., in scalar-tensor theories~\cite{Uzan:2010pm, Damour:1992kf, Damour:1993id, Anderson:2016aoi}.
We parametrize this time variation as $G \approx G_N\,[\,1 + G_1\,(t-t_0)\,]$, where $t_0$ is the present time and $G_N$ is the present-day value of the Newton constant.
Note that when computing the effect on a binary inspiral, keeping only the first-order term $G_1(t-t_0)\ll1$ is sufficient.
One can neglect the modification to the waveform amplitude, and consider only the dephasing contribution, which enters at $-4\,\mathrm{PN}$ order with respect to the leading-order quadrupole phase~\cite{tahura_parameterized_2018, Yunes:2009bv}:
\begin{equation}
	\delta\varphi = \Psi_\mathrm{TF2} \, C_{G_1} \frac{M}{M_\odot} \eta^{-1} \frac{G_1}{\mathrm{yr}^{-1}} (\pi M f)^{-8/3}\,,
	\label{eq:Gdot_dephasing}
\end{equation}
where $M$ is the total mass of the binary, $\eta= {m_1 m_2}/{M^2}$ is the symmetric mass ratio, $f$ is the wave frequency, $C_{G_1} = -3.52\times 10^{-15}$ and $\Psi_\mathrm{TF2} = (3/128\eta)(\pi M f)^{-5/3}$ is the leading-order phase of the TaylorF2 approximant.
As discussed in Sec.~\ref{sec:introduction}, the most stringent current bounds on $G_1$ come from quasi-static Solar-System experiments, $|G_1| \lesssim 10^{-12}$--$10^{-13}\,\mathrm{yr}^{-1}$~\cite{bambi_response_2005}, while strong-field GW observations from LVK comparable-mass binaries give limits several orders of magnitude weaker, $|G_1| \lesssim 10^{2}\,\mathrm{yr}^{-1}$~\cite{yunes_theoretical_2016}.

\subsection{Environmental dephasing: accretion and viscous migration}
\label{subsec:env}
About $10\%$ of observed supermassive black holes are active galactic nuclei (AGN)~\cite{Heckman:2014kza, Shankar:2007zg}, a duty cycle that we will use in Sec.~\ref{sec:hierarchical_inference} as order of magnitude for the active-fraction hyperparameter $\lambda$ of the population model.
Galaxy mergers generically drive gas inflows toward the galactic nucleus~\cite{Mihos:1995ri, Barnes:1996qt, Hopkins:2009td}, so MBHBs can plausibly be embedded in a gas-rich environment in which the surrounding material settles into a circumbinary disk~\cite{Mayer:2007vk}. This disk drives the accretion and migration effects discussed below.
In the presence of such a gaseous environment, the two black holes accrete and grow in mass.
The accretion timescale is expected to be comparable to the Salpeter timescale $\tau = 4.5\times 10^7~\mbox{yr}$~\cite{Salpeter:1964kb}, so that, assuming that both components accrete at the same rate, the total mass of the system evolves as $M(t) =M_0 \exp\!\left( f_\mathrm{Edd}(t-t_0)/\tau\right) \simeq M_0\,\bigl[1 + f_\mathrm{Edd}\,(t-t_0)/\tau\bigr]$, where $M_0$ is the total mass at $t_0$.
Since the LISA observation time is much shorter than $\tau$, we keep only the linear order.
Inserting this time-dependent mass into the quadrupole formula and imposing angular-momentum conservation~\cite{caputo_gravitational-wave_2020}, one finds that the induced dephasing is
\begin{equation}
	\delta\varphi = \Psi_\mathrm{TF2} \, C_\mathrm{acc} \frac{M}{M_\odot} \eta^{-1} \frac{f_\mathrm{Edd}}{0.1} (\pi M f)^{-8/3}\,,
	\label{eq:acc_dephasing}
\end{equation}
where $C_\mathrm{acc} = 1.75\times 10^{-23}$.

Another effect that can arise in a gaseous astrophysical environment is the migration torque exerted by the disk onto the binary.
In this case, which we will refer to as viscous migration, the dephasing enters at the same $-4\,\mathrm{PN}$ order~\cite{garg_imprint_2022, Garg:2024oeu}:
\begin{equation}
\delta\varphi = \Psi_\mathrm{TF2} \, C_\mathrm{mig} \frac{M}{M_\odot} \eta^{-2} \frac{\xi f_\mathrm{Edd}}{0.1} (\pi M f)^{-8/3},
\label{eq:mig_dephasing}
\end{equation}
where $C_\mathrm{mig}= 1.0\times 10^{-22}$ and $\xi$ encodes the efficiency of the disk-binary coupling; extrapolating the results of numerical simulations to thin disks, $\xi$ may reach values of $\mathcal{O}(10)$~\cite{Dittman:2022}. The extra factor $\eta^{-1}\xi$ in Eq.~\eqref{eq:mig_dephasing} relative to Eq.~\eqref{eq:acc_dephasing} makes migration more efficient than accretion at the same Eddington ratio, especially for asymmetric mass ratios.

\section{Population models and single-event analysis}
\label{sec:population_catalogs}

The parameter space spanned by LISA MBHBs covers several orders of magnitude in both chirp mass and distance, and the number of detected events in any given region of this space depends sensitively on the assumed astrophysical history of the MBHB population. We therefore rely on a suite of semi-analytic population models, each encoding different choices for the seeding mechanism, the delay time between the galaxy merger and the subsequent MBH merger, and the impact of supernova (SN) feedback on MBH accretion. The catalogs produced by these models are the basis for the detectability and bias estimates presented in the rest of this section.

\subsection{Population models}
\label{subsec:catalog_description}

The formation history of MBHBs is not well constrained observationally, and plausible evolutionary sequences spanning high redshift to the present yield substantially different catalog realizations. We adopt the semi-analytic model for the formation and evolution of galaxies and MBHs originally introduced in~\cite{Barausse:2012fy}. Additions and improvements to specific aspects of the model were subsequently introduced in~\cite{Sesana:2014bea, Antonini:2015sza, Bonetti:2017lnj, Bonetti:2018tpf, Barausse:2020mdt}. The catalogs are publicly available as part of the BACH data release~\cite{BACH2023}.

We consider five population models, three HS and two LS. In the HS class, the black-hole seeds arise from the collapse of proto-galactic disks induced by bar instabilities~\cite{Volonteri:2007ax}, with typical seed masses of the order of $\sim 10^{5}\,M_\odot$:
\begin{itemize}
    \item HS-nod-SN (from~\cite{Barausse:2020mdt}): no delay between the galaxy merger and the MBH merger (except for the dynamical-friction timescale, including tidal effects, between dark-matter halos), and including the effect of SN feedback on MBH accretion;
    \item HS-nod-SN-high-accr (from~\cite{Barausse:2020mdt, barausse_implications_2023}): as above, but with the MBH accretion rate boosted by a factor of $\sim 4$;
    \item Q3-nod (from~\cite{klein_science_2016}): an earlier HS realization with no delay and no SN feedback on accretion.
\end{itemize}
In the LS class, the seeds form instead from the remnants of population III stars in high-redshift, low-metallicity systems~\cite{Madau:2001sc}:
\begin{itemize}
    \item LS-nod-SN (from~\cite{Barausse:2020mdt}): no delay between the galaxy merger and the MBH merger (except for the dynamical-friction timescale, including tidal effects, between dark-matter halos), and including the effect of SN feedback on MBH accretion;
    \item popIII-d (from~\cite{klein_science_2016}): an earlier LS realization that includes, on top of the dynamical-friction timescale with tidal effects between dark-matter halos, further delay contributions from stellar hardening, triple-MBH interactions, and gas-driven migration.
\end{itemize}
Together, these five population models span a representative range of astrophysically motivated assumptions about the seeding mechanism and accretion history of MBHs across cosmic time.

\begin{figure*}
	\includegraphics[width =\textwidth]{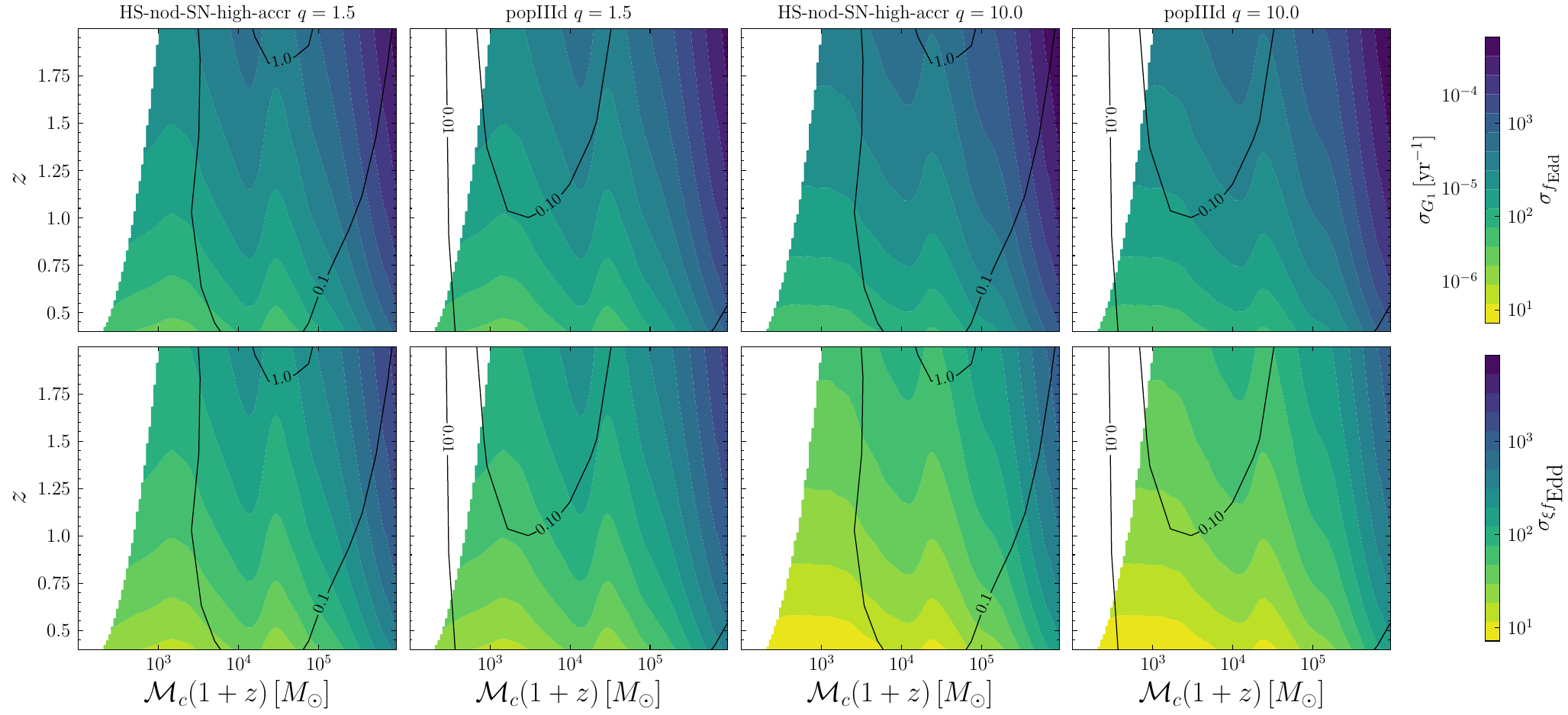}
	\caption{Sky-averaged $1\sigma$ uncertainty on the three $-4\,\mathrm{PN}$ parameters across the (detector-frame) chirp-mass--redshift plane, for $10\,\mathrm{yr}$ of LISA observation and mass ratios $q=1.5$ (left) and $q=10$ (right). Top panels: uncertainty on the Eddington fraction $f_\mathrm{Edd}$ (accretion) and on $G_1$ (modified gravity). Bottom panels: uncertainty on the combination $\xi f_\mathrm{Edd}$ relevant for viscous migration. Contours show the merger-rate density $\mathrm{d}^3 N / (\mathrm{d}\log_{10}\mathcal{M}_{c,\,z}\,\mathrm{d}z\,\mathrm{d}t_c)\times 1\,\mathrm{yr}$ for the HS-nod-SN-high-accr and popIII-d models, which bracket the population scenarios considered in this work. Points with sky-averaged $\mathrm{SNR}<8$ across the three TDI channels are masked out.}
    \label{fig:waterfall}
\end{figure*}

\subsection{LISA sensitivity across parameter space}
\label{subsec:Det_LISA}
We now compare the sensitivity of LISA to the effects described in Sec.~\ref{sec:pn4} across the parameter space.
Since all of these effects enter at the same PN order, we consider an agnostic dephasing term $\delta\varphi = \Psi_\mathrm{TF2} \, C_0 \, (\pi M f)^{-8/3}$, which allows one to estimate the LISA sensitivity to generic $-4\,\mathrm{PN}$ phase terms.
The model-specific dependence on the intrinsic and beyond-GR source parameters is encoded in the agnostic coefficient $C_0$, and we can recover information on $f_\mathrm{Edd}$ or $G_1$ via the physical mapping defined by Eqs.~\eqref{eq:acc_dephasing} and~\eqref{eq:mig_dephasing}.
Note that the Jacobian of this transformation is essentially constant over the posterior, so that a uniform prior on $C_0$ maps to an effectively uniform prior on $f_\mathrm{Edd}$ (or $G_1$).

To map out LISA's reach across parameter space, we evaluate the Fisher matrix on a grid over the plane of (detector-frame) chirp mass and redshift  $(\mathcal{M}_{c,\,z}, z)$, with $\mathcal{M}_c \equiv (m_1 m_2)^{3/5}/(m_1+m_2)^{1/5}$ and $\mathcal{M}_{c,z} = \mathcal{M}_c (1+z)$, at four representative values of the mass ratio $q \equiv m_1/m_2 \geq 1$: $q = 1.5,\, 3,\, 5,\, 10$.
For each point in the $(\mathcal{M}_{c,z},\,z)$ grid we compute the Fisher matrix
\begin{equation}
	\Gamma_{ij} = \braket{\partial_i h(\theta) \lvert \partial_j h(\theta)}
	\label{eq:fisher}
\end{equation}
over $500$ Monte Carlo realizations of the extrinsic parameters (sky position, inclination, and polarization angle), while fixing the mass ratio and spins to fiducial values.
The nominal LISA mission duration is $4\,\mathrm{yr}$, with an extended-phase option reaching $10\,\mathrm{yr}$~\cite{colpi_lisa_2024}; throughout this work we adopt the optimistic $10\,\mathrm{yr}$ figure. 
For accretion and viscous migration under realistic astrophysical assumptions, MBHBs in the LISA band have $C_0 \sim 10^{-15}$.
Since the dephasing depends linearly on $C_0$, and this additional contribution is strongly subdominant compared to the radiation-reaction-driven phase evolution, the Fisher matrix is approximately independent of $C_0$.

In Fig.~\ref{fig:waterfall} we show the sky-averaged statistical error on the physical parameter characterizing each of the three $-4\,\mathrm{PN}$ channels, for mass ratios $q=1.5$ and $q=10$, together with the merger-rate distribution of two of the population models discussed in Sec.~\ref{subsec:catalog_description}. We mask out the points for which the sky-averaged signal-to-noise ratio (SNR) across the three LISA Time-Delay Interferometry (TDI) channels is below $8$.
The top panels show the $1\sigma$ uncertainty on the Eddington fraction $f_\mathrm{Edd}$ (accretion) and on the fractional drift rate $G_1$ (modified gravity); the bottom panels show the corresponding uncertainty on the combination $\xi f_\mathrm{Edd}$ relevant for viscous migration.
The figures are overlaid with the merger-rate $\frac{\mathrm{d}^3 N}{\mathrm{d}\log_{10} \mathcal{M}_{c,\,z}\mathrm{d}z\mathrm{d}t_c}\times 1~\mbox{yr}$ contours for the HS-nod-SN-high-accr and popIII-d models, which bracket the range of formation models explored above. 
In agreement with previous single-event forecasts~\cite{garg_imprint_2022}, the sensitivity to all three $-4\,\mathrm{PN}$ deformations is concentrated at low chirp mass and low redshift, where the early inspiral spends months to years in band; conversely, the massive sources that dominate the HS-nod-SN-high-accr merger rate fall in the poorly-constrained top-right corner of each panel.
Even in the optimal cases---long observation time and low redshift and mass---the expected average statistical error on the environmental parameters is generally well above $f_\mathrm{Edd}=1$ for both the accretion and the viscous-migration case.

Naively one might expect LS models to be the optimal case for the inference of these negative-PN effects. We find instead that the HS models deliver more mergers per year with low statistical error on the $-4\,\mathrm{PN}$ physical parameters.
The underlying reason is the difference in seed mass between the two classes.
LS mergers typically have chirp masses in the range $500$--$1000\,M_\odot$, with the merger---where most of the SNR is accumulated---outside the LISA band; despite their long in-band inspiral, such events are therefore faint and only the closest to us are detectable. SN feedback~\cite{habouzit_blossoms_2017} further reduces the detectable population by stunting seed growth~\cite{Barausse:2020mdt, habouzit_blossoms_2017}, so that the merger rate at $\mathcal{M}_c \sim 10^{4}$--$10^{5}\,M_\odot$ is strongly suppressed in LS-nod-SN relative to a popIII-remnant scenario without SN feedback.
In the HS models, by contrast, the seeds are already born at $\sim 10^{5}\,M_\odot$, i.e.\ directly in the LISA chirp-mass window, so the merger rate at $\mathcal{M}_c \sim 10^{4}$--$10^{5}\,M_\odot$ is largely insensitive to the SN-feedback prescription and is typically an order of magnitude higher than in LS-nod-SN (Fig.~\ref{fig:waterfall}).

\subsection{Single-event inference}
\label{subsec:specific_realization}
The Fisher-matrix parameter-estimation results of Fig.~\ref{fig:waterfall} describe the sky-averaged reach of LISA as a function of the intrinsic source parameters. To move from this sky-averaged picture to more realistic per-catalog forecasts, we now draw specific realizations of each MBHB population and analyze them event by event.
For each population model the intrinsic parameters (component masses, spins, and luminosity distance) are drawn from the associated merger-rate distribution, while the extrinsic parameters (sky position, time and phase of coalescence, polarization, and inclination) are drawn from uninformative (uniform or isotropic) distributions.
We retain only events exceeding the detection threshold  $\mbox{SNR} > 8$.

\begin{figure*}
    \centering
    \includegraphics[width =0.8\textwidth]{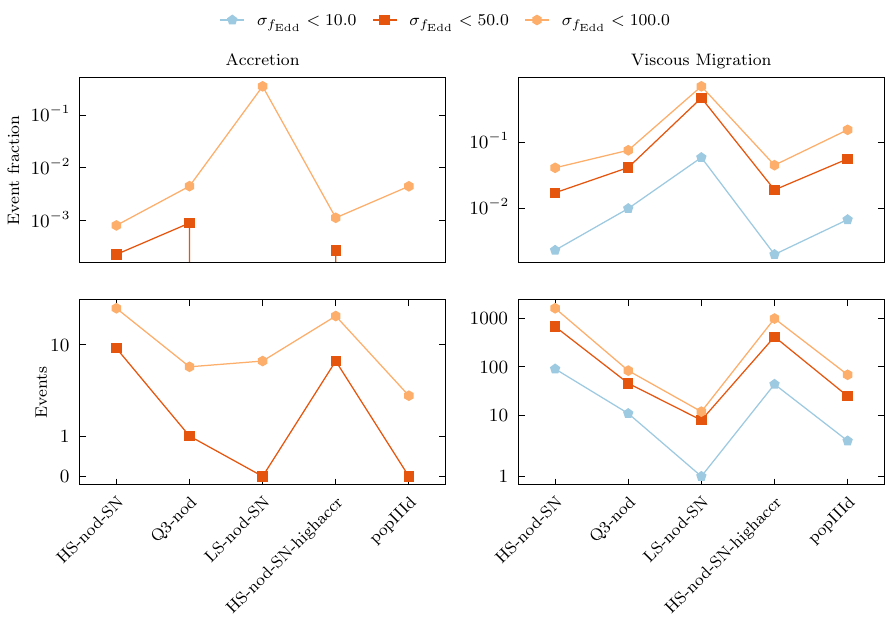}
    \caption{Number and fraction of events detectable in $10\,\mathrm{yr}$ of LISA observation whose Fisher-predicted single-event uncertainty $\sigma_{f_\mathrm{Edd}}$ falls below a threshold, for various population models. Top panels: fraction; bottom panels: absolute number. \emph{Left:} accretion injection, with thresholds $\sigma_{f_\mathrm{Edd}}<50$ (red) and $<100$ (orange). \emph{Right:} viscous-migration injection, with thresholds $\sigma_{f_\mathrm{Edd}}<10$ (blue), $<50$ (red) and $<100$ (orange).}
    \label{fig:accretion_events}
\end{figure*}

We analyze each catalog under two contrasting assumptions for how the parameter estimation treats potential $-4\,\mathrm{PN}$ contributions to the signal. In the \emph{first scenario}, the waveform templates include a free $-4\,\mathrm{PN}$ phase deformation parametrized by $C_0$, whose posterior we estimate from the Fisher information matrix computed over the intrinsic parameters \textit{and} $C_0$. 
Figure~\ref{fig:accretion_events} summarizes the outcome for the accretion and viscous-migration templates, respectively: for each $10\,\mathrm{yr}$-catalog, we report the number and fraction of detectable events with $\sigma_{f_\mathrm{Edd}}$ below $50,\,100$ (accretion) or $10,\,50,\,100$ (migration).
The tighter threshold is not reported for the accretion case, because in most catalogs no event satisfies $\sigma_{f_\mathrm{Edd}}\lesssim 10$.

In a second scenario, we assume that parameter estimation is performed with vacuum GR templates, and any unmodeled negative-PN contribution induces a systematic bias on the recovered parameters. We quantify the per-event bias with the Cutler--Vallisneri formula~\cite{Cutler2007},
\begin{equation}
    \Delta \theta_j = \Gamma_{ij}^{-1}\,\left\langle d - h(\theta)\,\middle|\,\partial_j h(\theta)\right\rangle\,,
    \label{eq:cutvallisneri}
\end{equation}
where $\Gamma_{ij}$ is now the vacuum-template Fisher matrix, evaluated at the at the true value of the parameters. The residual between the data and the vacuum template can be approximated as
\begin{equation}
    d - h(\theta) = h(\theta)\bigl(e^{i\delta\varphi} -1\bigr) \simeq i\,\delta\varphi\,h(\theta)\,,
    \label{eq:approx_residual}
\end{equation}
which, inserted into Eq.~\eqref{eq:cutvallisneri}, shows that the bias scales linearly with $f_\mathrm{Edd}$ (or with $G_1$). It is therefore sufficient to compute the bias at a fiducial $f_\mathrm{Edd}$ or $G_1$, and to rescale to the desired value. We have verified this by comparing the bias computed with the approximated and with the full residual formulae.

The dominant effect of neglecting a negative-PN contribution is a shift in the recovered chirp mass, while the recovered spins and mass ratio remain essentially unaffected with respect to the statistical errors, even for super-Eddington injections. Moreover, the chirp-mass shift itself is negligible for the vast majority of events, consistent with the limited detectability of generic $-4\,\mathrm{PN}$ corrections across the parameter space discussed in Sec.~\ref{subsec:Det_LISA}. To quantify the shift, we normalize it to the Fisher-predicted $1\sigma$ uncertainty on $\mathcal{M}_c$,
\begin{equation}
    \frac{\delta\mathcal{M}_c}{f_\mathrm{Edd}} \;\equiv\; \frac{\Delta \mathcal{M}_c}{f_\mathrm{Edd}\,\sigma_{\mathcal{M}_c}}\,,
\end{equation}
and count, for each   $10\,\mathrm{yr}$- catalog, the number of events for which this normalized shift exceeds $0.01$, $0.1$, and $1$ (assuming at $f_\mathrm{Edd}=1$). The results are reported in Table~\ref{tab:acc_biases}. No event in any catalog is shifted by more than $10\%$ of the statistical uncertainty: accretion at physically plausible rates leaves essentially no imprint on the vacuum-template chirp-mass measurement, and only at unrealistically high accretion rates, $f_\mathrm{Edd}\sim 10^{2}$--$10^{3}$, does the shift become comparable to the statistical error.

\begin{table}[htp]
    \centering
    \begin{tabular}{c|ccc}
    \toprule
    catalog & $\frac{\delta \mathcal{M}_c}{f_\mathrm{Edd}} > 0.01$ & $\frac{\delta \mathcal{M}_c}{f_\mathrm{Edd}} > 0.1$ \\
    \hline
    HS-nod-SN-high-accr & 33 & 0 \\
    HS-nod-SN & 41 & 0 \\
    LS-nod-SN & 11 & 0 \\
    popIII-d  & 15 & 0 \\
    Q3-nod    & 5  & 0 \\
    \bottomrule
    \end{tabular}

    \caption{Average number of sources observed in $10\,\mathrm{yr}$ whose vacuum-template chirp-mass bias, normalized to the Fisher $1\sigma$ uncertainty, exceeds the thresholds $\Delta\mathcal{M}_c/(f_\mathrm{Edd}\,\sigma_{\mathcal{M}_c}) > 0.01,\,0.1$, assuming every event undergoes accretion at $f_\mathrm{Edd}=1$, for various population models.}
    \label{tab:acc_biases}
\end{table}

Repeating the same exercise for a viscous-migration injection at $f_\mathrm{Edd} = 1$, we find a substantially stronger effect (Table~\ref{tab:mig_biases}). This is expected: the migration dephasing scales as $\eta^{-2}$ rather than the $\eta^{-1}$ of accretion [Eqs.~\eqref{eq:acc_dephasing}--\eqref{eq:mig_dephasing}], and is therefore amplified for asymmetric mass ratios. Already at $f_\mathrm{Edd}=1$, individual sources in several catalogs show a chirp-mass shift comparable to the statistical uncertainty. Viscous migration can be further enhanced or suppressed by the internal structure of the disk through the efficiency factor $\xi$: current numerical simulations give $\xi=\mathcal{O}(1)$~\cite{Duffell_2020}, but those simulations do not resolve realistically thin disks because of their computational cost, and it has been argued that for a thin circumbinary disk $\xi$ may be as large as $\mathcal{O}(10)$~\cite{Garg:2024oeu, Dittman:2022}. Under this more optimistic assumption, Table~\ref{tab:mig_biases} implies that up to a few events per year in the HS catalogs could exhibit chirp-mass shifts at or above the statistical error.

\begin{table}[htp]
    \centering
    \begin{tabular}{c|ccc}
    \toprule
    catalog & $\frac{\delta \mathcal{M}_c}{f_\mathrm{Edd}} > 0.01$ & $\frac{\delta \mathcal{M}_c}{f_\mathrm{Edd}} > 0.1$ & $\frac{\delta \mathcal{M}_c}{f_\mathrm{Edd}}>1$ \\
    \hline
    HS-nod-SN-high-accr & 1132 & 66 & 1 \\
    HS-nod-SN & 1812 & 117 & 2 \\
    LS-nod-SN & 15 & 3 & 1 \\
    popIII-d & 139 & 5 & 0 \\
    Q3-nod & 89 & 11 & 0 \\
    \bottomrule
    \end{tabular}

    \caption{Same as Table~\ref{tab:acc_biases}, but for a viscous-migration injection at $f_\mathrm{Edd}=1$ and fiducial $\xi=1$.}
    \label{tab:mig_biases}
\end{table}

Together, these per-event uncertainties and biases on the template parameters are the input to the hierarchical population analysis of Sec.~\ref{sec:hierarchical_inference}, where they are combined to yield joint population constraints on $\dot G$ and on the environmental parameters.

\section{Population-level inference}
\label{sec:hierarchical_inference}
For single events, the precision to which a negative-PN contribution can be measured is limited.
The different physical effects considered in this work leave essentially the same imprint: they all induce a $-4\,\mathrm{PN}$ dephasing and are therefore degenerate at the level of an individual source.
Information at the \emph{population} level, however, can break part of this degeneracy. A putative drift of Newton's constant is expected to act in the same way on every source, whereas environmental effects act only on the fraction of binaries embedded in a gaseous environment, with disk properties that vary from event to event~\cite{Jones_2016}.

Combining many noisy single-event measurements into a joint constraint on the population-level hyperparameters is naturally handled by hierarchical Bayesian inference, which we briefly review below before applying it to our problem.
\subsection{Hierarchical Bayesian framework}
\label{subsec:hierarchical_refresher}
We consider a set of $N$ independent events with data $\mathbf{D}=\{D_1,D_2,\ldots,D_N\}$.
The full set of parameters is split into the single-event parameters $\boldsymbol{\theta} = \{\theta_1,\,\theta_2,\,\dots,\,\theta_N\}$ where $\theta_i$ is the set of parameters for the event $i$, and the population-level hyper-parameters $\boldsymbol{\Lambda}$, the latter characterizing the probability distribution of $\boldsymbol{\theta}$ across the population.
The joint posterior then reads
\begin{equation}
	p(\boldsymbol{\theta},\boldsymbol{\Lambda}\,|\,\mathbf{D}) \propto p(\mathbf{D}\,|\,\boldsymbol{\theta})\,p(\boldsymbol{\theta}\,|\,\boldsymbol{\Lambda})\,p(\boldsymbol{\Lambda}),
		\label{eq:hyper-posterior}
\end{equation}
and marginalizing over the single-event parameters yields the posterior on the hyper-parameters.

Directly inferring the full posterior is pretty involved, especially for an high-dimensional posterior.
In situations where the posteriors $p(D_i|\theta_i)$ are uncorrelated with each other the hyperposterior can be obtained throgh importance reweighting~\cite{Thrane_Talbot_2019}:
\begin{eqnarray}
	p(\mathbf{D}|\boldsymbol{\Lambda})&= \prod_{i} \int \mathrm{d}\theta_i\, p(D_i|\theta_i) p(\theta_i|\boldsymbol{\Lambda})\nonumber \\
	&=\prod_{i} \int \mathrm{d}\theta_i\, \dfrac{p(\theta_i|D^i)p(D_i)}{p(\theta_i)} p(\theta_i|\boldsymbol{\Lambda}) \nonumber \\
&\simeq \prod_i p(D_i)\mathbb{E}\bigg[\dfrac{p(\theta_i|\boldsymbol{\Lambda})}{p(\theta_i)}\bigg]
	\label{eq:hyplike}
\end{eqnarray}
where the index $i$ labels the different events in the catalog, $p(\boldsymbol{\theta}_i)$ are the agnostic priors (independent of the hyperparameters), $p(D_i)$ is the evidence for the single event computed with these agnostic priors, and
\begin{equation}
	\mathbb{E}[f(\theta_i)] = \frac{1}{N_s}\sum_{\theta_i\in p(\theta_i|D_i)} f(\theta_i)
\end{equation}
where $N_s$ is the number of samples extracted from the posterior $p(\theta_i|D_i)$.

Because the hyper-parameters control the underlying distribution of events, a naive combination of the single-event likelihoods would overweight high-SNR signals and largely ignore marginally detectable ones.
To avoid this bias, the hyper-posterior must be normalized by a selection factor, which measures the fraction of parameter space hosting detectable sources for a given choice of $\boldsymbol{\Lambda}$~\cite{Mandel:2018mve,Fishbach_2018,Sadiq:2024xsz}.
This correction is essential when inferring the mass or redshift distribution of the population, since the SNR depends strongly on these parameters.:
In our case, however, we are not interested in the binary distribution itself but only in the deviation from vacuum GR. Since all the effects considered here enter only through a small dephasing with negligible impact on the SNR, the selection function is effectively constant across the hyper-parameter space and can be dropped when computing Bayes factors.

\subsection{Population constraints on $\dot G$}
\label{subsec:Gdot_population}
For a time-dependent Newton constant, we can assume that the physical parameter $G_1$ is common to all sources.
As such, the informed prior takes the following form:
\begin{equation}
	p(\boldsymbol{\theta}| G_1) = \prod_{i} ^N  \delta\bigg(G_1 - \frac{C_{0,\,i} \eta_i}{C_G M_i}\bigg)
\end{equation}
where we use the index $i$ to specify the event in the catalog.
Inserting this into Eq.~\eqref{eq:hyper-posterior}, the delta function in the informed prior fixes $C_0^i = (C_G M_i / \eta_i)\,G_1$ for each event, so that each event contributes a Gaussian posterior on $G_1^i$ once the vacuum parameters are marginalized over. The hyperposterior on the universal $G_1$ is then the product of these single-event Gaussians, and is therefore Gaussian by construction with variance
\begin{equation}
	\sigma_{G_1} = \bigg(\sum_{i}^N \Gamma_{G_1 G_1}^i -\Gamma_{k\,G_1}^i C_{kj}^i\,\Gamma_{j\,G_1}^i\bigg)^{-1/2}\,,
	\label{eq:SigmaG1}
\end{equation}
where the Latin indices $k,\,j$ run over the vacuum parameters, the index $i$ labels the event in the catalog, and $C^i$ is the inverse of the vacuum-only Fisher submatrix for event $i$.
Equivalently, the term in parentheses is the inverse-variance sum $\sum_i 1/\sigma_{G_1,i}^2$, where $\sigma_{G_1,i}$ is the marginal single-event uncertainty on $G_1$, obtained by inverting the full per-event Fisher matrix over $(\theta_\mathrm{vac}^i, G_1)$ and reading off the $(G_1, G_1)$ diagonal of the resulting covariance matrix.

\begin{figure}
    \centering
    \includegraphics[width=\columnwidth]{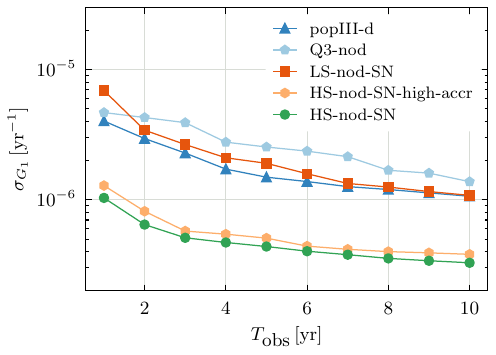}
    \caption{Population-level statistical error $\sigma_{G_1}$ on the time derivative of Newton's constant, as a function of the LISA mission duration.
 For the LS-nod-SN we report the mean across the different catalogues.}
    \label{fig:Gdot_mes}
\end{figure}

In Fig.~\ref{fig:Gdot_mes} we report the constraint on $G_1$ for different population models and observation times; in the bottom panel we compare the different models. Since the detection rate for the LS-nod-SN model is $\mathcal{O}(1~\mbox{yr}^{-1})$, the total statistical error depends strongly on the specific catalog realization.
In the plot we report the average statistical error on $G_1$ across 19 catalogue realizations for the LS-nod-NS model.
For the remaining models, we report only one catalogue realization, as the results do not change significantly among realizations.

In the HS models and in the popIII-d model, the improvement in resolution after the first few years of observation is smooth, indicating that the rate at which new events contribute appreciably to the hyperparameter constraint is roughly constant.
In none of the population models considered does the statistical error on $G_1$ drop below $\mathcal{O}(10^{-7}\,\mathrm{yr}^{-1})$.
While this is significantly better than the constraints obtained with LVK, it is a few orders of magnitude weaker than what can be probed with other long-lived sources in the LISA band (SOBHBs, EMRIs).

Note that, at linear order in $G_1$, the uncertainty $\sigma_{G_1}$ is independent of the fiducial value of $G$ itself: the leading-order correction to any gravitational-wave observable enters as $\delta O \propto G_1\, t$, with no dependence on the absolute normalization of $G$. This facilitates comparison between different hypotheses.

As the posteriors are Gaussian, it is then straightforward to compute the evidence for the modified-gravity model as
\begin{equation}
	p(\mathbf{D}) = \frac{\sqrt{2\pi}\sigma_{G_1}}{\mbox{V}_{G_1}} e^{-\mathcal{L}\textsubscript{max}}\prod_{i}  \frac{\sqrt{(2\pi)^d\det(C^i)}}{\mbox{V}\textsubscript{vac}}
	\label{eq:evidence_Gdot}
\end{equation}
where $d$ is the number of vacuum parameters, $\mbox{V}\textsubscript{vac}$ is the prior volume for the vacuum parameters, $\mbox{V}_{G_1}$ is the prior volume of $G_1$, and:
\begin{equation}
	\mathcal{L}\textsubscript{max} = \sum_{i} \frac{\tilde{\Gamma}^i G_{1\, i}^{2}}{2} - \sigma_{G_1}^2\frac{(\sum_{i} \tilde{\Gamma}^i G_{1\,i} )^2}{4} 
\end{equation}
with $\tilde{\Gamma} =  \Gamma_{G_1 G_1}- \Gamma_{k\,G_1} C_{kj}\,\Gamma_{j\,G_1}$, while $G_{1\,i}$ is the true value of $G_1$ for each event in the catalog.
We validated the analytical result of Eq.~\eqref{eq:evidence_Gdot} against a numerical computation based on Eq.~\eqref{eq:hyplike} and a nested-sampling algorithm~\cite{Skilling:2006gxv}; the prior on $G_1$ is specified in Sec.~\ref{subsec:astrophysical_inference}.

\subsection{Population constraints on environmental effects}
\label{subsec:astrophysical_inference}
For an astrophysical population, the value of $f_\mathrm{Edd}$ is expected to vary from source to source.
We assume that the Eddington ratio follows a power-law distribution with fixed upper cutoff $f_\mathrm{max} = 10$ and, in line with the AGN duty cycle quoted in Sec.~\ref{subsec:env}, that only a fraction of binaries are active.
We therefore introduce two population-level hyperparameters: the active fraction $\lambda$ and the power-law slope $\alpha$,
\begin{equation}
	p(f_\mathrm{Edd}\,|\,\lambda,\alpha) = \lambda\, \alpha\, \frac{f_\mathrm{Edd}^{\alpha - 1}}{f_\mathrm{max}^\alpha} + (1-\lambda)\, \delta(f_\mathrm{Edd}).
    \label{eq:population_model}
\end{equation}

We consider three recovery hypotheses:
\begin{itemize}
	\item $\mathcal{H}_V$: a vacuum template, with no further population-level assumption;
	\item $\mathcal{H}_E$: a $-4\,\mathrm{PN}$ template for the single events and, at the population level, a fraction $\lambda$ of binaries undergoing viscous migration while the rest merge in vacuum;
	\item $\mathcal{H}_{\dot G}$: a $-4\,\mathrm{PN}$ template for the single events and, at the population level, a universal time-dependent Newton constant.
\end{itemize}
Under $\mathcal{H}_E$ we adopt uniform priors on $\lambda$ in $[0,1]$ and on $\alpha$ in $[0.01, 2]$; under $\mathcal{H}_{\dot G}$ we adopt a uniform prior on $G_1$ in $[-5\times 10^{-6}, 5\times 10^{-6}]\,\mathrm{yr}^{-1}$.
To compare two recovery hypotheses $\mathcal{H}_A$ and $\mathcal{H}_B$ we compute the Bayes factor
\begin{equation}
	\mathcal{B}^A_B = \frac{p(\mathbf{D}|\mathcal{H}_A)}{p(\mathbf{D}|\mathcal{H}_B)},
	\label{eq:bayes}
\end{equation}
so that $\mathcal{B}^A_B > 1$ favors $\mathcal{H}_A$ over $\mathcal{H}_B$.
For $\mathcal{H}_V$ and $\mathcal{H}_{\dot G}$ the evidences can be computed analytically, giving
\begin{equation}
	\bayes{V}{\dot G} = \exp\bigg(\frac{\sigma_{G_1}^2 \big(\sum_i \tilde{\Gamma}^i G_{1\,i}\big)^2  }{4}\bigg)\frac{\sqrt{2\pi}\sigma_{G_1}}{V_{G_1}}.
	\label{eq:Bayes_vacG}
\end{equation}
For $\mathcal{H}_E$, instead, the evidence does not admit a closed-form expression and is evaluated numerically with the nested-sampling algorithm implemented in \texttt{dynesty}~\cite{Speagle:2019ivv}.

To speed up the likelihood evaluation under the environmental hypothesis we retain only events with $\sigma_{f_\mathrm{Edd}}<20$.
The discarded events contribute the same multiplicative factor to the evidence of both $\mathcal{H}_V$ and $\mathcal{H}_E$ within our priors, and therefore drop out of the Bayes factor.
After this cut, no catalog retains usable events in the pure-accretion case, so we drop that comparison; for the viscous-migration injections most catalogs are left with fewer than $10$ events over $10\,\mathrm{yr}$ of observation, and we restrict the analysis to the HS models, where the count is larger.

\begin{figure}[htpb]
	\centering
	\includegraphics[width=\linewidth]{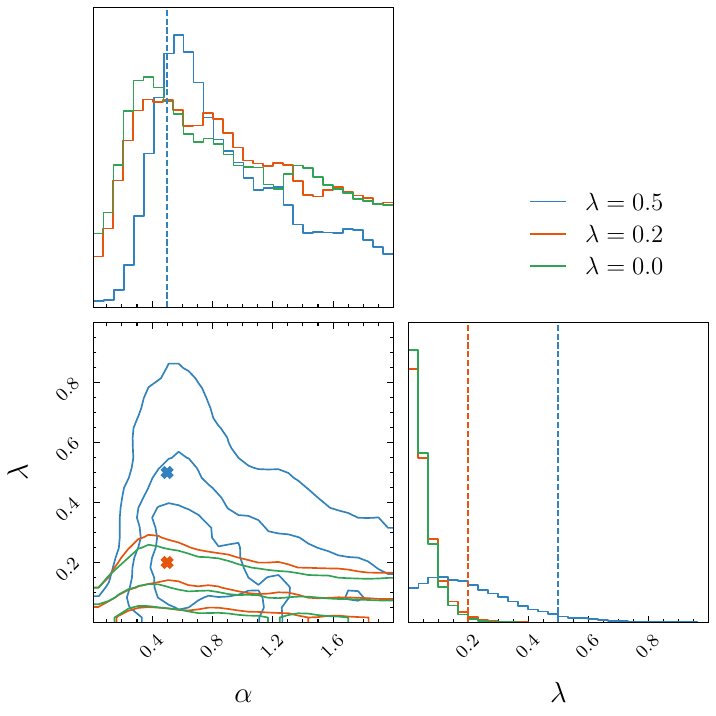}
	\caption{Population-level posteriors on the hyperparameters $(\lambda,\alpha)$ for three injections analyzed under the environmental hypothesis: a vacuum injection (green), and two viscous-migration injections with $\alpha=0.5$ and active fractions $\lambda=0.2$ (red) and $\lambda=0.5$ (blue). In all three cases the recovered posteriors are fully compatible with vacuum.}
	\label{fig:env_posterior}
\end{figure}

Figure~\ref{fig:env_posterior} compares the recovered $(\lambda,\alpha)$ posteriors for three injections: a vacuum injection, and two viscous-migration injections with $\alpha=0.5$ and active fractions $\lambda=0.2$ and $\lambda=0.5$.
For the $\lambda=0.2$ injection the posterior is essentially indistinguishable from the vacuum-injection one.
The $\lambda=0.5$ posterior visibly shifts away from vacuum, and the injected value lies well outside the posterior recovered from the vacuum injection.
Even in this most favorable case, however, the posteriors are not informative enough to pin down the astrophysical properties of the population: we recover only an upper limit on the active fraction $\lambda$ and a $2\sigma$ lower limit on the slope $\alpha$, and all posteriors are fully compatible with vacuum.
Consistent with this, Table~\ref{tab:Bayes} --- which reports the pairwise base-10 logarithms of the Bayes factors among $\mathcal{H}_V$, $\mathcal{H}_E$ and $\mathcal{H}_{\dot G}$ for the six injections considered --- favors $\mathcal{H}_V$ over $\mathcal{H}_E$ for both environmental injections. Increasing the active fraction from $\lambda=0.2$ to $\lambda=0.5$ shrinks $|\log_{10}\bayes{V}{E}|$ as expected, but the additional prior volume associated with the environmental hyperparameters still acts as an Occam's-razor penalty that the data are not informative enough to overcome.

\begin{figure}[htpb]
	\centering
	\includegraphics[width=\columnwidth]{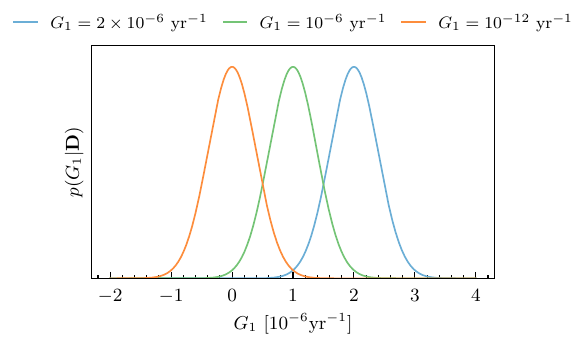}
	\caption{Population-level posteriors on $G_1$ for three modified-GR injections: $G_1 = 2\times10^{-6}\,\mathrm{yr}^{-1}$, $G_1 = 10^{-6}\,\mathrm{yr}^{-1}$, and $G_1 = 10^{-12}\,\mathrm{yr}^{-1}$ (in blue, compatible with Solar-System constraints). The Solar-System-compatible case is effectively centered at $G_1=0$.}
	\label{fig:Gdot_post}
\end{figure}

\begin{table}[htpb]
	\centering
	\caption{Base-10 logarithms of Bayes factors between the vacuum ($\mathcal{H}_V$), environmental ($\mathcal{H}_E$) and modified-gravity ($\mathcal{H}_{\dot G}$) hypotheses, for the six representative injections discussed in the text. Positive values favor the numerator hypothesis.}
	\label{tab:Bayes}

	\begin{tabular}{cccc}
		\toprule
		injection	 & $\log_{10} \bayes{V}{E}$ & $\log_{10} \bayes{\dot G}{E}$ & $\log_{10} \bayes{V}{\dot G}$\\
		 \midrule
Migration $(\lambda=0.5)$ & $-0.46$ & $0.07$ & $-0.52$\\
Migration $(\lambda=0.2)$ & $-1.71$ & $0.04$ & $-1.75$\\
Modified GR $(G_1 = 2\times10^{-6}~\mbox{yr}^{-1})$ & $-1.79$ & $-2.77$ & $0.97$\\
Modified GR $(G_1 = 10^{-6}~\mbox{yr}^{-1})$ & $-1.78$ & $-0.63$ & $-1.15$\\
Modified GR $(G_1 = 10^{-12}~\mbox{yr}^{-1})$ & $-1.73$ & $0.13$ & $-1.85$\\
Vacuum & $-1.73$ & $0.13$ & $-1.84$\\
		 \bottomrule
		
	\end{tabular}
\end{table}
We repeat the analysis for modified-GR injections; the posteriors are shown in Fig.~\ref{fig:Gdot_post}.
Since the statistical error on $G_1$ is $\mathcal{O}(10^{-7}\,\mathrm{yr}^{-1})$, an injection compatible with the Solar-System bound, $G_1 = 10^{-12}\,\mathrm{yr}^{-1}$, is indistinguishable from pure GR.
We therefore consider larger injections that still respect the LVK bound on $G_1$.
From Eq.~\eqref{eq:Bayes_vacG}, the modified-gravity Bayes factor scales as $\exp(G_1^2/\sigma_{G_1}^2)$, so doubling the injected $G_1$ from $10^{-6}\,\mathrm{yr}^{-1}$ to $2\times10^{-6}\,\mathrm{yr}^{-1}$ raises $\bayes{\dot G}{V}$ by roughly two orders of magnitude.
Only the $G_1 = 2\times 10^{-6}\,\mathrm{yr}^{-1}$ case yields a clear preference for the modified-gravity hypothesis.

We now compare $\mathcal{H}_E$ and $\mathcal{H}_{\dot G}$ directly. When the data are well described by vacuum, the marginalized evidence under $\mathcal{H}_E$ is dominated by the region near $\lambda=0$, where the environmental model reduces to $\mathcal{H}_V$; the integral therefore equals the vacuum evidence multiplied by an Occam factor (the ratio of the posterior width near $\lambda=0$ to the $(\lambda,\alpha)$ prior volume). The same happens for $\mathcal{H}_{\dot G}$ near $G_1=0$. For our priors the two Occam factors are numerically similar, so the Bayes factor between $\mathcal{H}_E$ and $\mathcal{H}_{\dot G}$ stays close to unity --- as confirmed by the $\log_{10}\bayes{\dot G}{E}$ column of Table~\ref{tab:Bayes}, which gives $|\log_{10}\bayes{\dot G}{E}|\leq 0.13$ for the four vacuum-compatible injections. The approximation breaks for the two strong modified-gravity injections ($G_1=10^{-6}$ and $2\times10^{-6}\,\mathrm{yr}^{-1}$), where the evidence of $\mathcal{H}_{\dot G}$ is genuinely enhanced over vacuum and $\log_{10}\bayes{\dot G}{E}$ drops to $-0.63$ and $-2.77$, respectively.

As a final cross-check, we fit the environmental hypothesis $\mathcal{H}_E$ to data generated under modified-GR injections. The recovered $(\lambda,\alpha)$ posteriors are fully consistent with $\lambda=0$, so the environmental recovery does not falsely claim a viscous-migration signal even when the data contain a real $\dot G$ signal. Table~\ref{tab:Bayes} bears this out from two angles: the $\log_{10}\bayes{V}{E}$ column is essentially unchanged across the three modified-GR injections (it stays near $-1.75$), confirming that the V-vs-E balance does not respond to the true $G_1$; in parallel, $\log_{10}\bayes{V}{\dot G}$ moves from $-1.85$ at $G_1=10^{-12}\,\mathrm{yr}^{-1}$ to $+0.97$ at $G_1=2\times10^{-6}\,\mathrm{yr}^{-1}$, showing that the $\dot G$ analysis correctly responds to the injection strength.

\section{Conclusion}
In this work we investigated whether the astrophysical environment of LISA MBHBs could leave a detectable imprint at the population level, and whether it could mimic---and thereby bias---LISA's tests of GR. We focused on three $-4\,\mathrm{PN}$ deviations from vacuum templates: gas accretion onto the components, viscous migration in a circumbinary disk, and a secular drift of Newton's constant. All three are formally degenerate for a single source, but the degeneracy is broken at the population level: a $\dot G$ drift is universal across the catalog, whereas the environmental effects act only on a fraction of binaries and with a magnitude that varies from event to event.

Across the five population models considered, the most informative sources are low-mass MBHBs at low redshift, which are rare irrespective of the seeding model. Even at the optimistic LISA $10\,\mathrm{yr}$ mission duration, no event is individually sensitive to accretion at realistic Eddington ratios: at $f_\mathrm{Edd}=1$, vacuum-template recovery shifts no event's chirp mass by more than $10\%$ of the statistical error. Viscous migration is a more promising channel because of its stronger $\eta^{-2}$ mass-ratio scaling: up to $\sim 8$ events per year in the HS models admit a single-event measurement with $\sigma_{f_\mathrm{Edd}}<10$ at fiducial $\xi=1$, and at the optimistic thin-disk value $\xi\sim 10$ a few HS-catalog events per year can exhibit a vacuum-template chirp-mass bias at or above the statistical error. This handful of relatively noisy detections is, however, not enough to drive a population-level preference for an environmental signature.

At the population level we compared three recovery hypotheses: vacuum, a hierarchical environmental model parameterized by the active fraction $\lambda$ and the power-law slope $\alpha$ of the Eddington-ratio distribution, and a universal time-dependent Newton constant. For $\dot G$, the population-level statistical error settles at $\sigma_{G_1}\sim 10^{-7}\,\mathrm{yr}^{-1}$ across all the catalogs considered---significantly better than current bounds from ground-based GW observations, but several orders of magnitude weaker than the Solar-System limit and than what can be reached with longer-lived LISA sources such as EMRIs and SOBHBs. For the environmental hypothesis, all the injections considered---including a viscous-migration injection with active fraction $\lambda=0.5$ and a super-Eddington Eddington-ratio tail ($f_\mathrm{max}=10$)---yield population posteriors that are fully compatible with vacuum, although they do produce a non-trivial upper limit on $\lambda$ and a $2\sigma$ lower bound on $\alpha$. A marked Bayes-factor preference for the modified-gravity hypothesis emerges only for injected $G_1$ several orders of magnitude above the Solar-System limit; such values are not yet excluded in the strong-field, dynamical regime probed by ground-based GW observations, but no realistic theoretical motivation singles them out.

Taken together, these results indicate that, under astrophysically realistic assumptions, environmental contamination of LISA's tests of GR with MBHBs is negligible---and, symmetrically, that a population-wide detection of an environmental imprint in LISA MBHBs is unlikely in any of the population scenarios we considered. One important caveat is in order: our analysis is restricted to circular inspirals, while recent works~\cite{Takatsy:2025bfk, Zwick:2025qzv} have shown that environmental dephasing can be substantially amplified in eccentric orbits. Extending the present population-level framework to eccentric MBHBs would simultaneously sharpen the prospects for an environmental detection and the corresponding risk of $\dot G$-test bias, and we leave this investigation to future work.
\begin{acknowledgments}
We acknowledge support from the European Union's Horizon ERC Synergy Grant ``Making Sense of the Unexpected in the Gravitational-Wave Sky'' (Grant No.\ GWSky--101167314). This work has been supported by the Agenzia Spaziale Italiana (ASI), Project n.~2024-36-HH.0, ``Attivit\`a per la fase B2/C della missione LISA''.
This work makes use of  \texttt{numpy}~\cite{harrisArrayProgrammingNumPy2020a}, \texttt{matplotlib}~\cite{matplotlib} and \texttt{lisabeta}~\cite{lisabeta}.
\end{acknowledgments}

\bibliographystyle{apsrev}
\bibliography{references}
\end{document}